
\magnification \magstep1
\raggedbottom
\openup 4\jot
\voffset6truemm
\headline={\ifnum\pageno=1\hfill\else
\hfill{\it Quantum effects in Friedmann-Robertson-Walker
cosmologies}\hfill \fi}
\def\cstok#1{\leavevmode\thinspace\hbox{\vrule\vtop{\vbox{\hrule\kern1pt
\hbox{\vphantom{\tt/}\thinspace{\tt#1}\thinspace}}
\kern1pt\hrule}\vrule}\thinspace}
\rightline {DSF preprint 95/6, revised version}
\centerline {\bf QUANTUM EFFECTS IN}
\centerline {\bf FRIEDMANN-ROBERTSON-WALKER COSMOLOGIES}
\vskip 0.3cm
\centerline {\bf Giampiero Esposito,
Gennaro Miele, Luigi Rosa, Pietro Santorelli}
\vskip 0.3cm
\centerline {\it Istituto Nazionale di Fisica Nucleare, Sezione di Napoli}
\centerline {\it Mostra d'Oltremare Padiglione 20, 80125 Napoli, Italy;}
\centerline {\it Dipartimento di Scienze Fisiche}
\centerline {\it Mostra d'Oltremare Padiglione 19, 80125 Napoli, Italy.}
\vskip 0.3cm
\noindent
{\bf Abstract.} Electrodynamics for self-interacting scalar fields in
spatially flat Friedmann-Robertson-Walker space-times is studied. The
corresponding one-loop field equation for the expectation value of the
complex scalar field in the conformal vacuum is derived.
For exponentially expanding universes, the equations for the
Bogoliubov coefficients describing the coupling of the scalar
field to gravity are solved numerically. They yield a non-local
correction to the Coleman-Weinberg effective potential which
does not modify the pattern of minima found in static de
Sitter space. Such a correction contains a dissipative term which,
accounting for the decay of the classical configuration
in scalar field quanta, may be relevant for the reheating stage.
The physical meaning of the non-local term in the semiclassical field
equation is investigated by evaluating this contribution for various
background field configurations.
\vskip 0.3cm
\leftline {PACS numbers: 0260, 0370, 0420, 1115, 9880}
\vskip 100cm
\leftline {\bf 1. Introduction}
\vskip 1cm
\noindent
In a recent series of papers [1-3], some of the authors have studied
the one-loop effective potential for grand unified theories in de
Sitter space. Our main results were a better understanding of the
symmetry-breaking pattern first found in [4], a numerical approach to
small perturbations of de Sitter cosmologies [2], and the analysis of
SO(10) GUT theories in de Sitter cosmologies [3]. However, a constant
Higgs field, with de Sitter four-space as a background in the
corresponding one-loop effective potential, is only a mathematical
idealization. A more realistic description of the early universe is
instead obtained on considering a dynamical space-time such as the one
occurring in Friedmann-Robertson-Walker (hereafter referred to as FRW)
models.
Indeed, our early work appearing in [2] tried to study the case of
varying Higgs field by introducing a function of the Euclidean-time
coordinate which reduces to the 4-sphere radius of de Sitter in the
limit of constant Higgs field. Although the approximations made in [2]
were legitimate for numerical purposes, the gravitational part of the
action, and the one-loop effective potential, were not actually
appropriate for studying a dynamical cosmological model.

Thus, relying on the work in [5-6], this paper studies the first step
towards the completion of our programme, i.e. scalar electrodynamics
with a self-interaction term for the complex scalar field. The
semiclassical field equations in time-variable backgrounds contain
non-local terms, which describe the coupling of the scalar field to
the gravitational background. The analysis of these equations is
relevant for the reheating mechanism in inflationary cosmology and for
the dynamics of dissipation via particle production [7]. Our analysis
deals with Lorentzian space-time manifolds with FRW symmetries, and
does not rely on zeta-function regularization. Hence our geometric
framework is substantially different from the Riemannian
four-manifolds studied in [1-4], where the metric was
positive-definite. Section 2 derives the field equations for a class
of cosmological models where scalar electrodynamics is studied in FRW
universes. The Coleman-Weinberg potential, and its correction deriving
from the non-local term involving the Bogoliubov coefficients for the
coupling of the scalar field to gravity, are obtained numerically in
section 3. The semiclassical field equation and the physical meaning
of such non-local term are studied in detail in section 4.
Concluding remarks are presented
in section 5, and relevant details are given in the appendix.
\vskip 1cm
\leftline {\bf 2. Model and field equations}
\vskip 1cm
\noindent
For the reasons described in the introduction, we consider a complex
scalar field $\phi$ with a mass and a self-interaction term, coupled
to the electromagnetic potential $A_{\mu}$ in curved space-time. Hence
the action functional is
$$
I \equiv \int {\cal L} \; \sqrt{-{\rm det} \; g}
\; d^{4}x \; + \; {\rm boundary} \; {\rm terms}
\eqno (2.1)
$$
where (cf [6])
$$ \eqalignno{
{\cal L} & \equiv \Bigr[(\nabla_{\mu}+ie A_{\mu})
\phi^{\dagger}\Bigr]
\Bigr[(\nabla^{\mu}-ie A^{\mu})\phi \Bigr]
-m^{2} \phi^{\dagger} \phi
-{1\over 4 !}\lambda (\phi^{\dagger} \phi)^{2} \cr
&-\xi R \phi^{\dagger} \phi
-{1\over 4} F_{\mu \nu} F^{\mu \nu}.
&(2.2)\cr}
$$
With a standard notation, $\nabla$ is the Levi-Civita connection on
the background space-time, $\xi$ is a dimensionless parameter, $R$
denotes the trace of the Ricci tensor, and $F_{\mu \nu} \equiv
\nabla_{\nu}A_{\mu}-\nabla_{\mu}A_{\nu}$ is the electromagnetic-field
tensor. Boundary terms are necessary to obtain a well-posed
variational problem, and their form is obtained after integration by
parts in the volume integral in (2.1). Covariant derivatives of the
scalar field are here used to achieve a uniform notation (cf [6] and
[8]).

Following [5], we now split the complex scalar field $\phi$ as the sum
of a variable, real-valued background field $\phi_{c}$, and of a
complex-valued fluctuation $\varphi$, i.e.
$$
\phi=\phi_{c}+\varphi.
\eqno (2.3)
$$
The {\it conformal vacuum} [9] is here chosen, and the quantum
fluctuation $\varphi$ has vanishing expectation value in such a state,
$< \varphi > = 0$, so that $< \phi > = \phi_{c}$. The field equations
for $A_{\mu}, \phi_{c}$ and $\varphi$ are obtained by setting to zero
the corresponding functional derivatives of the action. As far as the
gauge potential $A_{\mu}$ is concerned, it is convenient to impose the
Lorentz gauge $\nabla^{\mu}A_{\mu}=0$. At this stage, to quantize the
theory, one can follow the Gupta-Bleuler method, or the Faddeev-Popov
procedure, or to eliminate the residual gauge freedom by imposing the
relativistic gauge condition proposed by Ford in [6], i.e.
$X^{\mu}A_{\mu}=0$, where $X$ is a timelike vector field. Such a field
admits a {\it natural} form in FRW space-times (see below). Thus, on
defining the operator $\cstok{\ } \equiv g^{\mu \nu} \nabla_{\mu}
\nabla_{\nu}$, the resulting form of the field equations is (cf [6])
$$
\Bigr(g^{\mu \nu}\cstok{\ } - R^{\mu \nu}\Bigr)A_{\nu}
=-2e^{2} A^{\mu} \phi^{\dagger} \phi
-ie \Bigr(\phi^{\dagger} \nabla^{\mu} \phi
-\phi \nabla^{\mu} \phi^{\dagger}\Bigr)
\eqno (2.4)
$$
$$ \eqalignno{
\; & \biggr[\cstok{\ }+m^{2}+\xi R +{1\over 12}
\lambda \phi_{c}^{2}-e^{2}<A_{\mu}A^{\mu}>
+{1\over 24}\lambda <\varphi^{2}
+(\varphi^{\dagger})^{2}+4 \varphi \varphi^{\dagger}>
\biggr]\phi_{c} \cr
&=ie<A^{\mu}\nabla_{\mu}(\varphi-\varphi^{\dagger})>
&(2.5)\cr}
$$
$$
\biggr[\cstok{\ }+m^{2}+\xi R +{1 \over 6}
\lambda \phi_{c}^{2}\biggr]\varphi
+{1\over 12} \lambda \phi_{c}^{2} \; \varphi^{\dagger}=0.
\eqno (2.6)
$$
Note that equations (2.4)-(2.6) have been obtained by retaining in the
action (2.1) only terms quadratic in the fluctuations $\varphi$ and
$A_{\mu}$, and setting to zero all terms involving
$\nabla^{\nu}A_{\nu}$ and its covariant derivatives in the field
equations. The latter condition is sufficient to derive (2.4).
Moreover, we require that $\nabla^{\mu}\phi_{c}$ should be
proportional to $X^{\mu}$ [6]. Note also that the contribution of
$$
\biggr[\cstok{\ }+m^{2}+\xi R +{1\over 12}
\lambda \phi_{c}^{2}\biggr]\phi_{c}
$$
has been neglected in the course of deriving (2.6), since equation
(2.5) implies that such a contribution is of second order in the
quantum fluctuations. By taking the complex conjugate of equation
(2.6), and defining
$
\varphi \equiv (\varphi_{1}+i \varphi_{2})/\sqrt{2},
$
$
\lambda_{1} \equiv \lambda/2
$
and
$
\lambda_{2} \equiv \lambda/6 ,
$
the addition and subtraction of the resulting equations leads
to decoupled equations for the real and imaginary parts of
$\varphi$, i.e.
$$
\biggr[\cstok{\ }+m^{2}+\xi R
+{1\over 2} \lambda_{j}\phi_{c}^{2}\biggr]
\varphi_{j}=0
\; \; \; \; {\rm for} \; {\rm all} \; j=1,2.
\eqno (2.7)
$$
Moreover, by virtue of our particular gauge conditions, equation (2.5)
takes the form (see the appendix)
$$
\biggr[\cstok{\ }+m^{2}+\xi R
+{1\over 12} \lambda \phi_{c}^{2}
-e^{2}<A_{\mu}A^{\mu}>
+{1\over 4}\lambda_{1}<\varphi_{1}^{2}>
+{1\over 4}\lambda_{2}<\varphi_{2}^{2}>
\biggr]\phi_{c}=0.
\eqno (2.8)
$$
Interestingly, the effects of quantum fluctuations in
equation (2.8) reduce to a linear superposition
of the self-interaction term studied in [5] and of the
electromagnetic term studied in [6], without any
coupling. The term $<A_{\mu}A^{\mu}>$ is evaluated
on considering the integral equation equivalent to (2.4),
as shown in [6] and in the appendix. Such an analysis proves
that, in the case of a spatially flat FRW background, the
renormalized expectation value of $A_{\mu}A^{\mu}$
in the conformal vacuum can be written as [6]
$$
<A_{\mu}A^{\mu}>=B \; X^{\mu} X^{\nu} R_{\mu \nu}
\eqno (2.9)
$$
where $B=9.682 \cdot 10^{-3}$. In particular, in a de Sitter universe,
$<A_{\mu}A^{\mu}>=12BH^{2}$, where $H$ is the Hubble parameter [6].

Following [5], we can now write the set of equations leading
to the solution of (2.7) and (2.8) in spatially flat FRW
backgrounds. Without making any approximation, if one defines
($a$ being the cosmic scale factor)
$$
\tau \equiv \int_{t_{0}}^{t}{dy \over a(y)}
\eqno (2.10)
$$
and, for all $j=1,2$
$$
\Omega_{j,k}^{2} \equiv k^{2}+a^{2}(\tau)
\biggr[m^{2}+\Bigr(\xi-{1\over 6}\Bigr)R(\tau)
+{1\over 2}\lambda_{j} \phi_{c}^{2}(\tau)\biggr]
\eqno (2.11)
$$
the one-loop field equations resulting from (2.7) and (2.8)
are (cf (3.32) and (3.33) in [5])
$$ \eqalignno{
\; & {1\over a^{2}} {d^{2}\phi_{c}\over d\tau^{2}}
+ {2\over a^{3}}{da\over d\tau}
{d\phi_{c}\over d\tau}
-e^{2}<A_{\mu}A^{\mu}>\phi_{c}
+{\partial V_{\rm eff}\over \partial \phi_{c}} \cr
&+{\phi_{c}\over 4\pi^{2}a^{2}}
\sum_{j=1}^{2}{1\over 2}\lambda_{j}\int_{0}^{\infty}dk \;
k^{2}\Omega_{j,k}^{-1}\Bigr[s_{j,k}+{\rm Re} \; z_{j,k}\Bigr]=0
&(2.12)\cr}
$$
$$
{d\over d\tau}s_{j,k}= \left({d\over d\tau} \log \;
\Omega_{j,k}\right){\rm Re} \; z_{j,k}
\eqno (2.13)
$$
$$
{d\over d\tau}z_{j,k}= \left({d\over d\tau}
\log \; \Omega_{j,k}\right)
\Bigr(s_{j,k}+{1\over 2}\Bigr)
-2i \; \Omega_{j,k} \; z_{j,k}.
\eqno (2.14)
$$
The form of $\partial V_{\rm eff} / \partial \phi_{c}$
is given in the appendix.
Equations (2.13) and (2.14) are necessary to find solutions of
(2.7) by using Fourier-transform techniques and a suitable
change of coordinates, as shown in [5]. The initial conditions
for $s_{j,k}$ and $z_{j,k}$ are the ones appropriate for
the choice of conformal vacuum, i.e. [5]
$$
s_{j,k}(\tau=0)=0 \; \; \; \; \; \; \; \;
z_{j,k}(\tau=0)=0.
\eqno (2.15)
$$
Of course, the values
of $m,\xi$ and $\lambda_{j}$ should be now regarded as the
renormalized values of such parameters [5].

Note that the second line of equation (2.12) is a typical non-local
term, resulting from the self-interaction of the scalar field
(see (2.7)) and from its coupling to the geometric background.
Moreover, a dissipative term exists which is part of the
non-local correction to the Coleman-Weinberg potential, and it is due
to the decay processes of $\phi_c$ in scalar field quanta. By means of
such decays, energy is transferred from $\phi_c$ to the relativistic
degrees of freedom $\varphi$
(i.e. radiation). As it is well known, if this
release of energy is sufficiently strong, radiation becomes dominant
and hence the inflationary phase ends. This leads to the reheating
stage, which is as important as the exponential expansion for the
dynamics of the early universe.
\vskip 5cm
\leftline {\bf 3. Numerical evaluation of the one-loop
effective potential}
\vskip 1cm
\noindent
In this section, for the physically relevant case of an exponentially
expanding FRW universe, we compute the non-local term on the second
line of (2.12). After integration of such a term with respect to
$\phi_c$, the resulting expression is compared with the
Coleman-Weinberg potential $V_{\rm eff}$.

The numerical analysis has been performed for a conformally invariant
scalar field (hence $\xi=1/6$ and $m=0$), by solving equations (2.13)
and (2.14) with the help of the NAG routine D02BAF, with initial
conditions (2.15). We have fixed $\lambda=10^{-2}$, which ensures the
reliability of the perturbative approach, and the Hubble parameter $H
= 10^{-1}~M_{\rm PL}$, which is an intermediate choice between a
chaotic model ($H \simeq M_{\rm PL}$) and a GUT inflationary phase ($H
\simeq 10^{-4}~M_{\rm PL}$).

In the case of exponential expansion in the time variable $t$,
which implies (see (2.10))
$$
a(\tau) = {a(0)\over (1 - a(0)H \tau)}
\eqno (3.1)
$$
and for a classical field configuration independent of $\tau$,
$\phi_{c}=\phi_{c0}$, we express the second line of (2.12) as a
function of $\tau$ and $\phi_{c0}$. In figure 1 we plot the integral
with respect to $\phi_{c0}$ of the above quantity when the conformal
time $\tau$ varies between $0$ and $1/H$, which corresponds to a large
$e$-fold number, $\phi_{c0} \in [0,10~M_{\rm PL}]$, and $a(0)$ has
been set to 1. This configuration for $a(\tau)$ and fixed $\phi_{c0}$
corresponds to an exact de Sitter space, but unlike [1-4], with a
Lorentzian signature.
The non-local term describes an energy exchange between the
gravitational field and the field $\phi$ (see (2.3)). Such an exchange
may lead to dissipative or non-dissipative processes in the early
universe, depending on the sign of the non-local term, as it will be
clear from the analysis in section 4. For a constant background field
$\phi_c$, only a scale factor varying in time yields the above
non-local corrections to the Coleman-Weinberg potential $V_{\rm eff}$.

In figure 2, we plot the Coleman-Weinberg potential $V_{\rm eff}(\phi_c)$
(resulting from the integration of (A.7) for $\mu_1=\mu_2=M_{\rm PL}$
and $V_{\rm eff}(0)=V_0 = 3 M_{\rm PL}^2 H^2/(8 \pi)$)
for the same choice of parameters.
As one can see from figures 1 and 2, the contribution of non-local terms
to the one-loop effective potential is very small, and hence its effect
on the semiclassical equation of motion is negligible in a first
approximation (see section 4).
\vskip 1cm
\leftline {\bf 4. Semiclassical field equation and non-local effects}
\vskip 1cm
\noindent
The analysis of the previous section has shown that the effect of the
Bogoliubov coefficients on the Coleman-Weinberg potential does not
modify the pattern of minima found in static de Sitter space [8]. It
is therefore legitimate to study the equation (2.12) when, in a first
approximation, its second line is neglected, and then to use the
resulting solution to evaluate the non-local correction. Here, the
value of $V_{\rm eff}(0)$ is set equal to $V_0$ as above, and it
dominates the energy, as it occurs in a de Sitter universe,
for $\phi_c < M_{\rm PL}$.

On studying the limiting form of equation (2.12) when the second line
is neglected, it can be easily seen that the self-interaction term
(see (A.7)) plays a key role in obtaining a sensible physical model.
In other words, for vanishing $\lambda$, equation (2.12) admits a
run-away exact solution for $\phi_c$ in the form (denoting by $D_1$
and $D_2$ two integration constants)
$$
\phi_{c}=D_{1} \exp{\left(-\beta_{1}Ht\right)}
+D_{2} \exp{\left(-\beta_{2}Ht\right)}
\eqno (4.1)
$$
where $\beta_{1} \equiv \Bigr(3+\sqrt{9+48Be^{2}}\Bigr)/2$ and
$\beta_{2} \equiv \Bigr(3-\sqrt{9+48Be^{2}}\Bigr)/2$, and we have
re-expressed $\tau$ in terms of $t$ by means of (2.10). This behaviour
results from the particular form taken by the potential in (2.12), i.e.
$- 6 B H^2 e^2 \phi_c^2$, which is unbounded from below.

We thus study the equation given by the first line of (2.12) with
non-vanishing $\lambda$, i.e.
$$
{1 \over a^{2}} {d^{2}\phi_{c} \over d\tau^{2}} +
{2 \over a^{3}} {da \over d\tau}
{d\phi_{c} \over d\tau}
-e^{2} <A_{\mu} A^{\mu}> \phi_{c}
+ {\partial V_{\rm eff} \over \partial \phi_{c}} = 0
\eqno (4.2)
$$
subject to the initial conditions $\phi_c(0) = M_{\rm PL}$ and
$d\phi_c/d\tau (\tau = 0) = 0$. In figures 3 and 4, $\phi_c(\tau)$ and
$d\phi_c/d\tau$ are plotted. As one might have expected,
$\phi_c(\tau)$ decreases until it reaches the zero value, by virtue of
the nature of the potential (see figure 2), whilst the kinetic energy
increases.

We have then used the solution of equation (4.2) to evaluate the
non-local effects in our model. For this purpose, starting from the
definitions of energy density $\rho_{\phi_{c}}$ and pressure
$p_{\phi_{c}}$ of the background scalar field in a de Sitter universe,
i.e.
$$
\rho_{\phi_{c}} \equiv {1 \over 2} {{\dot\phi}_{c}}^2
+ V_{\rm eff} - 6Be^2H^2\phi_{c}^2
\eqno (4.3)
$$
$$
p_{\phi_{c}} \equiv {1 \over 2} {{\dot\phi}_{c}}^2 -
V_{\rm eff} + 6Be^2H^2\phi_{c}^2
\eqno (4.4)
$$
one gets from (2.12) the equation
$$
{d\rho_{\phi_c}\over d\tau} + 3 H a ( \rho_{\phi_c} + p_{\phi_c} )
= - \Theta(\phi_c(\tau),\tau){d\phi_c \over d\tau}.
\eqno (4.5)
$$
With our notation,
$\Theta$ corresponds to the whole second line of (2.12),
$$
\Theta(\phi_c(\tau),\tau) \equiv
{\phi_{c}\over 4\pi^{2}a^{2}}
\sum_{j=1}^{2}{1\over 2}\lambda_{j}\int_{0}^{\infty}dk \;
k^{2}\Omega_{j,k}^{-1}\Bigr[s_{j,k}+{\rm Re} \; z_{j,k}\Bigr].
\eqno (4.6)
$$
If the right hand side of (4.5) can be viewed as a dissipative term,
which implies that it always takes negative values and depends
quadratically on $d\phi_c/d\tau$, the Bianchi identity leads to the
following equation for the energy density $\rho_R$ of radiation:
$$
{d\rho_{R}\over d\tau}
+ 4Ha \; \rho_{R} = \Theta(\phi_c(\tau),\tau)
{d\phi_c \over d\tau}.
\eqno (4.7)
$$

In figure 5 we plot the right hand side of (4.5) corresponding to the
solution of equation (4.2). Note that its sign turns out to be
positive for the largest part of the $\tau$-interval,
and hence cannot actually lead to dissipative effects. A
naturally occurring question is how to interpret this lack of
dissipation. Indeed, in the adiabatic approximation studied in [5],
the coupling of the scalar field to a fermionic field by a Yukawa term
produces a vacuum energy loss rate proportional to $(H a \phi_c +
d\phi_c/d\tau)d\phi_c/d\tau$. In this last expression the second term
is clearly a dissipative effect, since it does not depend on the sign
of the velocity and in the formula reported in [5] it occurs with the
correct sign to represent an energy loss. By contrast, the first term
does depend on the sign of $d\phi_{c}/d\tau$ and in de Sitter, where
$H$ is constant, it only represents a further quantum correction to
the energy and pressure of the $\phi_c$ fluid.

These considerations seem to suggest that also in our case, where one
deals with decays of the classical field configuration into its
quanta, the non-local term in the semiclassical field equation may be
essentially a linear combination of $\phi_c$ and $d\phi_c/d\tau$. If
this property holds, one can expect that the evaluation of $(H a
\phi_c + d\phi_c/d\tau)d\phi_c/d\tau$ may indicate when dissipative
effects are likely to occur, depending on whether the first or the
second term of this linear combination is dominant. In other words,
the sign of the right hand side of (4.5) shown in figure 5 can be
understood by pointing out that for the solution of (4.2), shown in
figures 3 and 4, the ratio $(d\phi_c/d \tau)/Ha\phi_c$ is smaller
than 1 whilst $d \phi_{c}/d \tau$ is
negative, when $\tau \in [0,9.2 \; M_{\rm PL}^{-1}]$.
Note that, in the neighbourhood of $\tau=9.2 \; M_{\rm PL}^{-1}$,
$-\Theta \; d\phi_{c}/d \tau$ vanishes, and this corresponds
to the value of $\tau$ such that the linear combination
$(Ha\phi_{c}+ d\phi_{c}/d \tau)$ vanishes.

As a further check of the conjecture about the
functional dependence of the non-local term of equation (2.12) on
$\phi_c$ and $d\phi_c/d \tau$, one should also analyze other
situations where, unlike before,
the ratio $(d\phi_{c}/d \tau)/Ha\phi_{c}$
is larger than 1. This can be done for example
by taking as a trial function for $\phi_c(\tau)$ the linear
combination $\phi_c(\tau) = \phi_{c0} + \alpha \tau$, where $\alpha$
and $\phi_{c0}$ are two arbitrary parameters. The resulting
analysis, performed by varying $\alpha$ and $\phi_{c0}$ so as
to reproduce the conditions
$(d\phi_{c}/d \tau)/Ha\phi_{c} > 1$
or $(d\phi_{c}/d \tau)/Ha\phi_{c} < 1$, seems to confirm that,
in de Sitter, $\Theta(\phi_c(\tau),\tau)$ is indeed dominated by a term
proportional to the combination $(H a \phi_c + d\phi_c/d\tau)$.
\vskip 1cm
\leftline {\bf 5. Concluding remarks}
\vskip 1cm
\noindent
This paper has studied scalar electrodynamics in a spatially flat FRW
universe, by including a self-interaction term for a conformally
invariant scalar field (cf [5,6,10]). On imposing the Lorentz gauge
and the supplementary condition (A.4), the one-loop equation for the
expectation value of the complex scalar field in the conformal vacuum
shows a linear superposition of the self-interaction term of [5] and
of the electromagnetic term [6]. The results of our investigation
are thus as follows.

First, the numerical solution for the Bogoliubov coefficients in
(2.13) and (2.14), subject to the initial conditions (2.15), has been
obtained and used to find the correction to the Coleman-Weinberg
effective potential by integrating with respect to $\phi_c$ the second
line of (2.12). The pattern of minima in the effective potential is
not modified by the non-local term in (2.12). For the particular
values of parameters considered in our investigation, the non-local
corrections turn out to be several orders of magnitude smaller.

Second, the limiting form of equation (2.12), i.e. equation (4.2),
has been studied, and its numerical solution has been used to
evaluate non-local effects in our cosmological model. Such a
solution corresponds to a slow-roll dynamics. Interestingly,
the approximate calculation of the function $\Theta$ defined in
(4.6) shows that $\Theta$ does not lead necessarily to dissipative
effects in the early universe. Nevertheless, in a de Sitter model,
the right hand side of equation (4.5)
is very well approximated by the same
combination of $\phi_{c}$ and $d\phi_{c}/d\tau$ which results
from the adiabatic case studied in [5], where the coupling of a scalar
field to a fermionic field was instead considered.
More precisely, we have
found that, for various forms of $\phi_{c}(\tau)$, the right hand side
of (4.5) is essentially given by
$$
-A\Bigr(Ha \; \phi_{c}+{d\phi_{c}\over d\tau}\Bigr)
{d\phi_{c}\over d\tau}
$$
where $A$ is a positive function of $\tau$. The first term in round
brackets leads to a further quantum correction to the energy
density of the scalar field, at least when the function $A$ is
slowly varying. The second term is purely dissipative
(see the end of section 2).

One of the main motivations for studying non-local corrections
to the Coleman-Weinberg potential was their possible relevance
for the reheating of the early universe [7]. Our analysis confirms
that they cannot be simply re-expressed by a term of the kind
$\Gamma {d\phi_{c}/d\tau}$, as first found in [7], where a
different approach to the effective action is used with respect
to [5]. A non-trivial open problem is the numerical evaluation
of the function $A$ in our de Sitter model. Moreover, it appears
necessary to understand how the form of $A$ depends on the
specific choice of the background field $\phi_{c}(\tau)$.
Last, but not least, the whole analysis should be repeated
for FRW models which are not in a de Sitter phase.
\vskip 1cm
\leftline {\bf Acknowledgments}
\vskip 1cm
\noindent
We are grateful to Renato Musto for useful discussions
on scalar electrodynamics, and to the European Union
for partial support under the Human Capital and
Mobility Programme. Anonymous referees made comments
which led to a substantial improvement of the original
manuscript.
\vskip 1cm
\leftline {\bf Appendix}
\vskip 1cm
\noindent
Following [6], we denote by $D_{R}^{\mu \nu}(x,x')$ the photon
retarded Green's function, which satisfies the equation
$$
\Bigr( \delta^{\mu}_{\rho} \; \cstok{\ } \;
-\nabla^{\mu} \nabla_{\rho} - R_{\rho}^{\mu} \Bigr)_{x}
D_{R}^{\rho \nu}(x,x') = g^{\mu \nu} \;
\delta(x,x')/ \sqrt{-{\rm det} \; g}.
\eqno (A.1)
$$
Thus, by writing $A_{\rm in}^{\mu}(x)$ for the photon in-field,
the solution of (2.4) to order $e$, in the Lorentz gauge
$$
\nabla^{\mu}A_{\mu}=0
\eqno (A.2)
$$
is given by [6]
$$
A^{\mu} \cong A_{\rm in}^{\mu}-ie \int d^{4}x' \;
\sqrt{-{\rm det} \; g(x')} \;
D_{R}^{\mu \nu}(x,x')
\Bigr(\phi^{\dagger}\nabla_{\nu}\phi
-\phi \nabla_{\nu}\phi^{\dagger}\Bigr)_{\rm in}.
\eqno (A.3)
$$
The residual gauge freedom of the problem is dealt with by
imposing the additional condition [6]
$$
X^{\mu}A_{\mu}=0
\eqno (A.4)
$$
where $X$ is the same timelike vector field appearing in
(2.9). Equations (A.2)--(A.4) imply that, on inserting (A.3)
into the right-hand side of (2.5), the only non-trivial
contribution still vanishes after integration by parts,
since
$$
D_{R}^{\mu \nu} \; \nabla_{\nu}\phi_{c}=0
\; \; \; \; \; \; \; \;
\nabla_{\mu} \; D_{R}^{\mu \nu}=0.
\eqno (A.5)
$$

In the equation (2.12) for the expectation value $\phi_{c}$
of the scalar field $\phi$, the derivative with respect to
$\phi_{c}$ of the one-loop effective potential is given, in the
case of non-vanishing renormalized mass, by (cf (3.15) of [5])
$$ \eqalignno{
{\partial V_{\rm eff}\over \partial \phi_{c}}
&=m^{2}\phi_{c}+\xi R \phi_{c}
+ {1\over 12}\lambda \phi_{c}^{3}
-{1\over 96 \pi^{2}} \lambda \Bigr(\xi -{1\over 6}\Bigr)
R \phi_{c}-{5\over 2304 \pi^{2}} \lambda^{2}\phi_{c}^{3} \cr
&+{1\over 384 \pi^{2}}\lambda \biggr(m^{2}
+\Bigr(\xi -{1\over 6}\Bigr)R
+{1\over 12} \lambda \phi_{c}^{2}\biggr) \phi_{c}
\log \left | {m^{2}+(\xi-(1/6))R+(1/12)\lambda\phi_{c}^{2}
\over m^{2}} \right | \cr
&+{1\over 128 \pi^{2}}\lambda \biggr(m^{2}
+\Bigr(\xi -{1\over 6}\Bigr)R
+{1\over 4} \lambda \phi_{c}^{2}\biggr) \phi_{c}
\log \left | {m^{2}+(\xi-(1/6))R+(1/4)\lambda\phi_{c}^{2}
\over m^{2}} \right | &(A.6)\cr}
$$
and for $m=0$ by (cf (3.16) of [5])
$$ \eqalignno{
{\partial V_{\rm eff}\over \partial \phi_{c}}
&=\xi R \phi_{c}
+{1 \over 12}\lambda \phi_{c}^{3}
+ {1\over 4608 \pi^{2}} \lambda^{2}\phi_{c}^{3}
\biggr( \log \left | {(\xi-(1/6))R+(1/12)\lambda\phi_{c}^{2}
\over (\lambda/12) \mu_1^2}\right | - { 11 \over 3} \biggr) \cr
&+ {1\over 512 \pi^{2}} \lambda^{2}\phi_{c}^{3}
\biggr( \log \left | {(\xi-(1/6))R+(1/4)\lambda\phi_{c}^{2}
\over (\lambda/4) \mu_1^2}\right | - { 11 \over 3} \biggr) \cr
&+{1\over 384 \pi^{2}} \lambda \Bigr( \xi - {1 \over 6}\Bigr)R
\phi_{c} \biggr( \log \left | {(\xi-(1/6))R+(1/12)\lambda\phi_{c}^{2}
\over (\xi - (1/6)) \mu_2^2} \right | - 1 \biggr) \cr
&+{1\over 128 \pi^{2}} \lambda \Bigr( \xi - {1 \over 6}\Bigr)R
\phi_{c} \biggr( \log \left | {(\xi-(1/6))R+(1/4)\lambda\phi_{c}^{2}
\over (\xi - (1/6)) \mu_2^2} \right | - 1 \biggr) .
&(A.7)\cr}
$$
In agreement with the notation of section 2, $m,\xi$ and $\lambda$
are the renormalized values of our
parameters, and $\mu_1$ and $\mu_2$ are completely arbitrary
renormalization points. Our equations
(A.6) and (A.7) are obtained by
imposing the renormalization conditions (3.12a)--(3.12c) of [5]
and bearing in mind that the numerical coefficients in our
equation (2.8) differ from the ones occurring in equation (2.19)
of [5], since we study a complex scalar field.
\vskip 0.3cm
\leftline {\bf References}
\vskip 0.3cm
\noindent
\item {[1]}
Buccella F, Esposito G and Miele G 1992
{\it Class. Quantum Grav.} {\bf 9} 1499
\item {[2]}
Esposito G, Miele G and Rosa G 1993
{\it Class. Quantum Grav.} {\bf 10} 1285
\item {[3]}
Esposito G, Miele G and Rosa G 1994
{\it Class. Quantum Grav.} {\bf 11} 2031
\item {[4]}
Allen B 1985 {\it Ann. Phys.}, {\it NY} {\bf 161} 152
\item {[5]}
Ringwald A 1987 {\it Ann. Phys.}, {\it NY} {\bf 177} 129
\item {[6]}
Ford L H 1985 {\it Phys. Rev.} D {\bf 31} 704
\item {[7]}
Boyanovski D, de Vega H J, Holman R, Lee D S and Singh A
1995 {\it Phys. Rev.} D {\bf 51} 4419
\item {[8]}
Allen B 1983 {\it Nucl. Phys.} B {\bf 226} 228
\item {[9]}
Birrell N D and Davies P C W 1982 {\it Quantum Fields in
Curved Space} (Cambridge: Cambridge University Press)
\item {[10]}
Shore G M 1980 {\it Ann. Phys.}, {\it NY} {\bf 128} 376
\vskip 1cm
\leftline {Figure captions:}
\vskip 1cm
\noindent
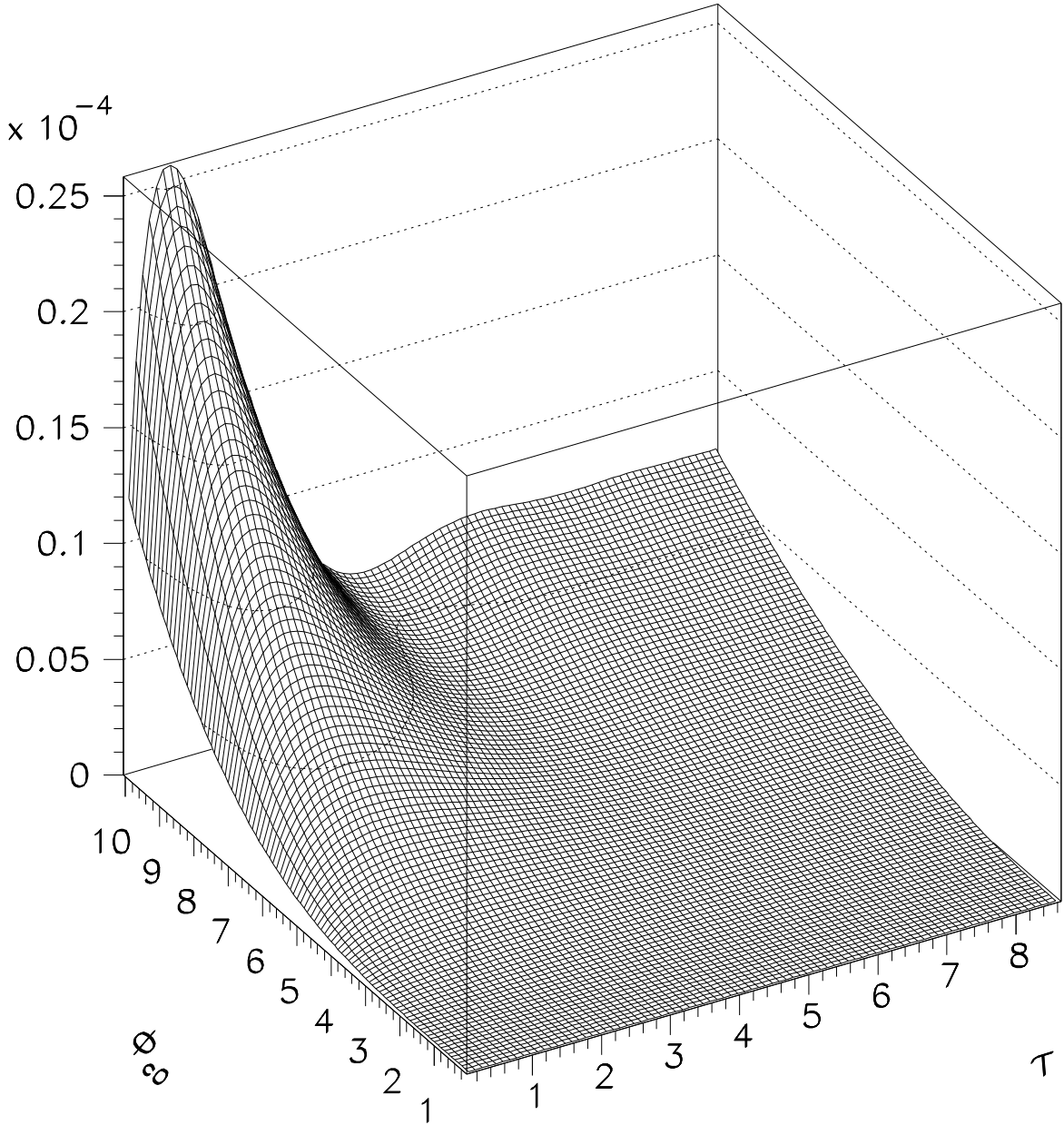
{\bf Figure 1.} In Planck units (used also in figures 2,3,4 and 5),
the correction
to the Coleman-Weinberg potential resulting from the
Bogoliubov coefficients in (2.12)--(2.14) is plotted
versus $\tau$ and $\phi_{c0}$.
\vskip 1cm
\noindent
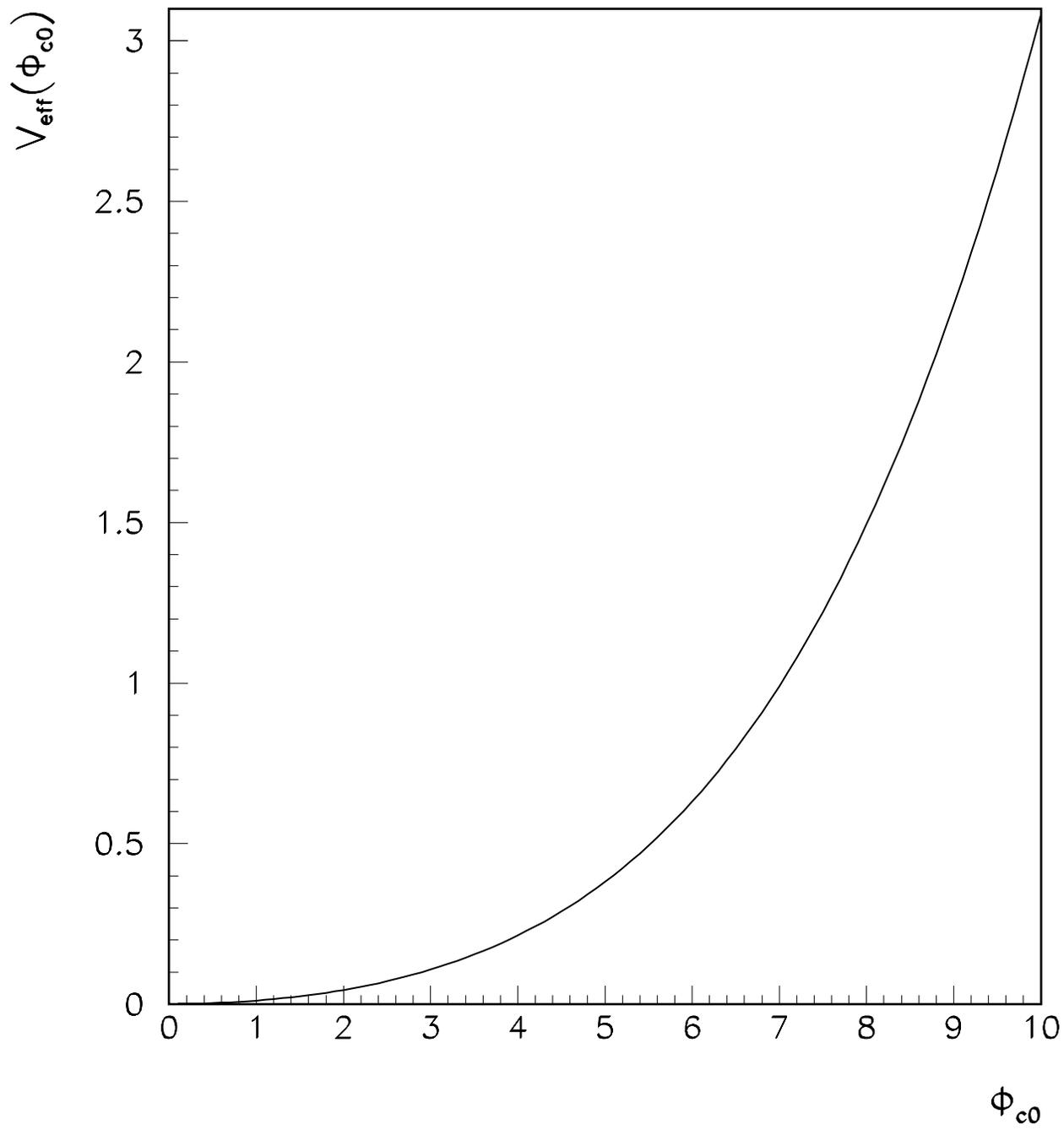
{\bf Figure 2.} The Coleman-Weinberg
one-loop effective potential
is plotted versus $\phi_{c0}$, in the case
of a de Sitter background.
\vskip 1cm
\noindent
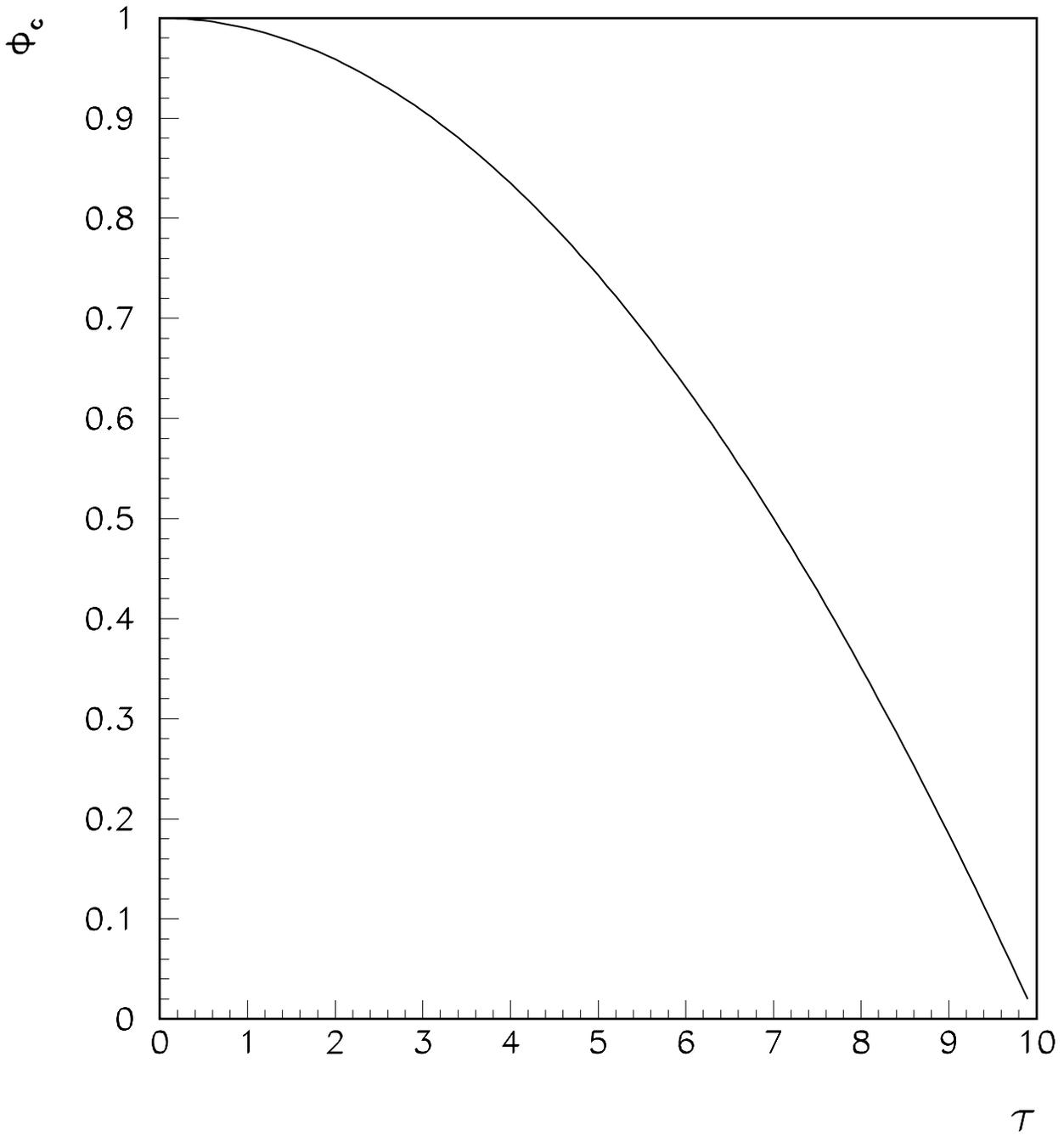
{\bf Figure 3.} The solution of the semiclassical field
equation (4.2) for the background field configuration
is plotted versus $\tau$.
\vskip 1cm
\noindent
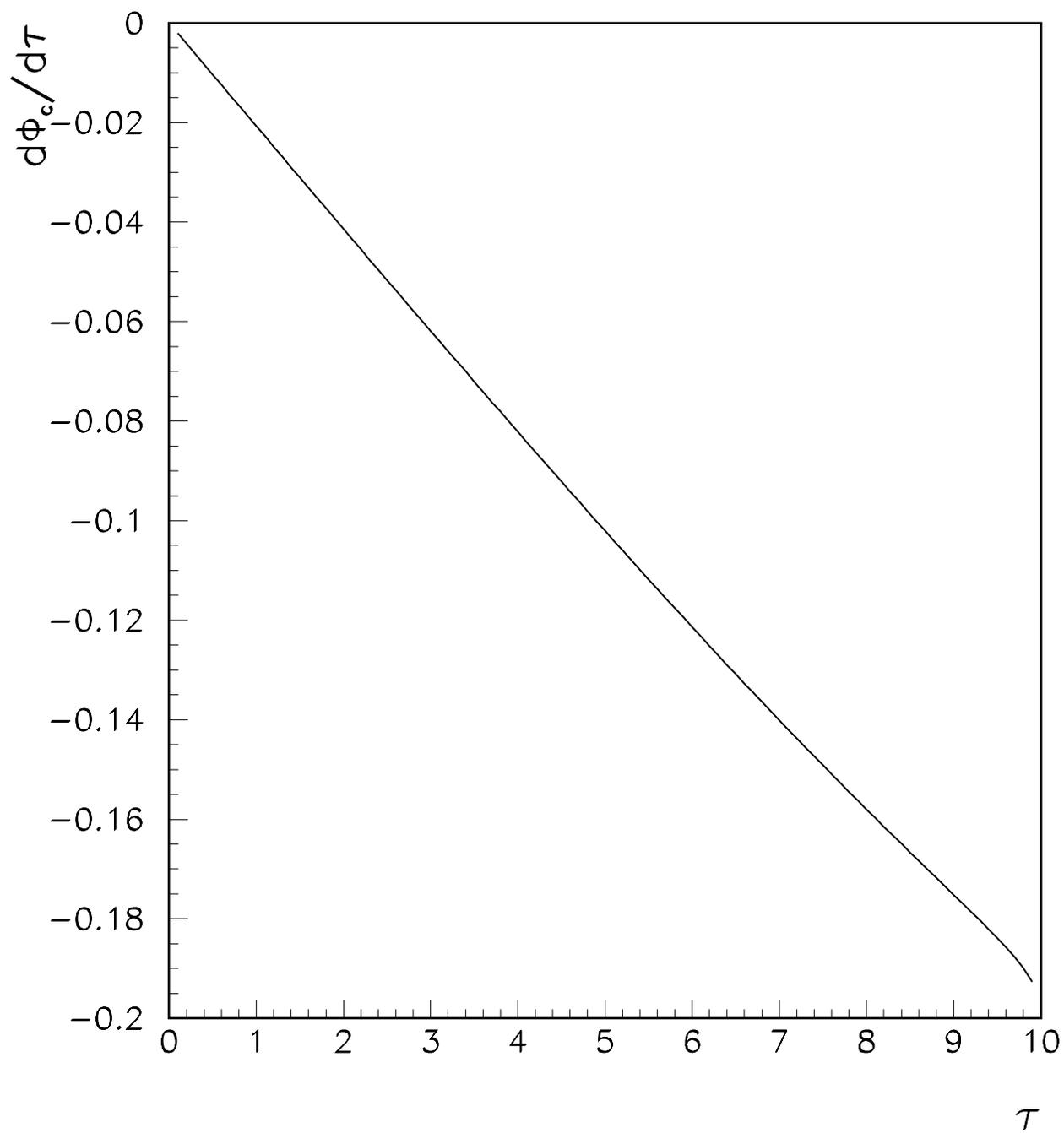
{\bf Figure 4.} For the numerical solution plotted in
figure 3, its derivative with respect to $\tau$ is
here shown.
\vskip 1cm
\noindent
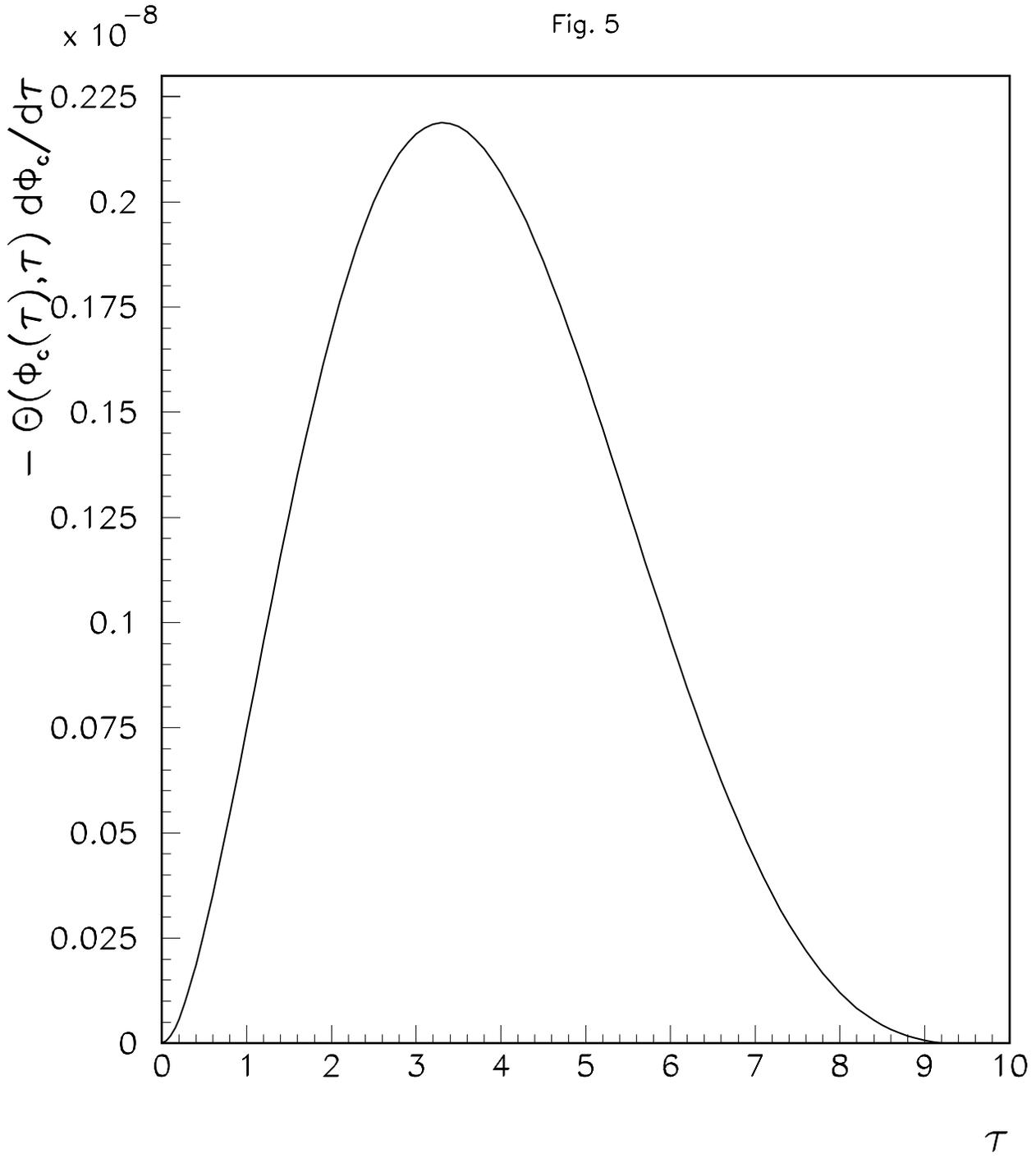
{\bf Figure 5.} The right hand side of equation (4.5) is
plotted versus $\tau$.

\bye